\documentclass[oribibl]{llncs}

\usepackage[utf8]{inputenc}
\usepackage[T1]{fontenc}
\usepackage{cite}
\usepackage{amsmath}
\usepackage{amssymb}
\usepackage{mathtools}
\usepackage{hyperref}
\usepackage{caption}
\usepackage{subcaption}
\usepackage{tikz}
\usepackage{pgfplots}
\usepackage{xspace}
\usepackage{drawmatrix}
\usepackage{float}
\usepackage{multirow}

\usepackage{aicescover}

\colorlet{graybg}{gray!10}
\colorlet{plot1}{red!75!white}
\colorlet{plot2}{green!75!black}
\colorlet{plot3}{blue}
\colorlet{plot4}{yellow!75!black}
\colorlet{plot5}{violet}
\colorlet{plot6}{cyan}
\colorlet{plot7}{orange}

\usetikzlibrary{
    external,
    calc,
    scopes,
}

\tikzset{
    external/mode=list and make,
    external/optimize command away=\aicescoverpage,
    external/export=false,
}
\tikzsetexternalprefix{figures/externalized/}

\pgfplotsset{
    every axis/.append style={
        axis background/.style={fill=graybg},
        legend cell align=left,
        xlabel near ticks,
        ylabel near ticks,
        enlarge x limits={value=0.05, auto},
        enlarge y limits={value=0.05, auto},
        width=\textwidth,
        height=.5\textwidth,
        ymajorgrids=true,
    },
    twocolplot/.style={
        height=.8\textwidth,
    },
    threecolplot/.style={
        height=.8\textwidth,
    },
    every axis plot/.append style={
        thick,
        line join=round
    },
}

\hypersetup{
    colorlinks=true,       
    linkcolor=blue!50!black,          
    citecolor=green!50!black,        
    filecolor={.5 0 0},      
    urlcolor=red!50!black,           
}

\newfloat{algorithms}{t}{loa}
\floatname{algorithms}{Algs.}
\DeclareCaptionSubType{algorithms}
\newfloat{algorithm}{t}{loa}
\floatname{algorithm}{Alg.}

\newcommand\dgeqrf{{\tt dgeqrf}\xspace}

\newcommand\dpotrf{{\tt dpotrf$_\texttt{L}$}\xspace}
\newcommand\dpotft{{\tt dpotf2$_\texttt{L}$}\xspace}
\newcommand\dgeqr{{\tt dgeqr2}\xspace}
\newcommand\dlarft{{\tt dlarft}\xspace}
\newcommand\dcopy{{\tt dcopy}\xspace}
\newcommand\dtrsm{{\tt dtrsm}\xspace}
\newcommand\dtrmm{{\tt dtrmm}\xspace}
\newcommand\dgemm{{\tt dgemm}\xspace}
\newcommand\dsyrk{{\tt dsyrk}\xspace}
\newcommand\dtrmmRLNU{\dtrmm{}$_\texttt{RLNU}$\xspace}
\newcommand\dgemmTN{\dgemm{}$_\texttt{TN}$\xspace}
\newcommand\dtrmmRUNN{\dtrmm{}$_\texttt{RUNN}$\xspace}
\newcommand\dgemmNT{\dgemm{}$_\texttt{NT}$\xspace}
\newcommand\dtrmmRLTU{\dtrmm{}$_\texttt{RLTU}$\xspace}
\newcommand\dtrsmLLN{\dtrsm{}$_\texttt{LLN}$\xspace}

\newcommand\blueball{%
    \tikz[baseline=(x.base)] \shade[ball color=plot3] circle (1.5pt) node (x) {\vphantom{fg}};\xspace%
}

\newcommand\parsum[1]{\textcolor{orange}{#1}}
\renewcommand\parsum[1]{}

\title{
    Cache-aware Performance Modeling and Prediction for Dense Linear Algebra
}
\author{
    Elmar Peise and Paolo Bientinesi
}
\institute{
    AICES, RWTH Aachen\\
    \email{\{peise,pauldj\}@aices.rwth-aachen.de}
}

\begin{document}
    \aicescoverpage
    \maketitle

    \begin{abstract}
    Countless applications cast their computational core in terms of dense
    linear algebra operations.  These operations can usually be implemented by
    combining the routines offered by standard linear algebra libraries such as
    BLAS and LAPACK, and typically each operation can be obtained in many
    alternative ways.  Interestingly, identifying the fastest
    implementation---without executing it---is a challenging task even for
    experts.  An equally challenging task is that of tuning each routine to
    performance-optimal configurations.  Indeed, the problem is so difficult
    that even the default values provided by the libraries are often
    considerably suboptimal; as a solution, normally one has to resort to
    executing and timing the routines, driven by some form of parameter search.
    In this paper, we discuss a methodology to solve both problems: identifying
    the best performing algorithm within a family of alternatives, and tuning
    algorithmic parameters for maximum performance; in both cases, we do not
    execute the algorithms themselves.  Instead, our methodology relies on
    timing and modeling the computational kernels underlying the algorithms, and
    on a technique for tracking the contents of the CPU cache.  In general, our
    performance predictions allow us to tune dense linear algebra algorithms
    within few percents from the best attainable results, thus allowing
    computational scientists and code developers alike to efficiently optimize
    their linear algebra routines and codes.
\end{abstract}

\parsum{orange: paragraph summary}


    \section{Introduction}
    \label{sec:intro}
    Most dense linear algebra (DLA) operations can be computed via multiple
alternative algorithms (``variants''), the performance of which can normally be
tuned by one or more configuration parameters.  Since the performance of these
algorithms depends on a variety of factors, including the problem size, the
target architecture, and the underlying libraries, selecting the optimal
combination of variant and configuration parameters becomes a real challenge.
However, such algorithms, including many covered by the Linear Algebra PACKage
(LAPACK)~\cite{lapack}, achieve their efficiency by building on a rather small
set of highly tuned kernels, most of which are provided by the Basic Linear
Algebra Subprograms (BLAS)~\cite{blas1, blas2, blas3}.  Given a linear algebra
algorithm $\mathcal A$ consisting of such kernels, we aim at predicting
$\mathcal A$'s performance, {\em without ever executing it}.  Our approach
relies on the typical layered structure of DLA libraries and on their
data-independent program flow;\footnote{%
  Eigensolvers are an exception, since their program flow always depends on the
  input data.} 
as a constraint, we only allow the execution of the kernels, and not of the
algorithm itself.

\parsum{approach}
In a nutshell, our approach consists of two stages: 1) generation of accurate
performance models for the computational building blocks (kernels) and, to
effectively use those models, 2) tracking the state of the CPU's cache
throughout the target algorithm.  In this paper, we extend and integrate our
work on automated performance modeling~\cite{modeling}, which only considered
in-cache operations, with that on cache tracking~\cite{caching}, in order to
construct reliable performance predictions for a spectrum of problem sizes,
ranging from small problems fitting in cache, to large ones that can only reside
in main memory.  We show that highly accurate time predictions for dense linear
algebra algorithms are attainable, and that such estimates are so close to the
observed performance, that it becomes possible not only to correctly rank the
algorithms, but to also tune their implementation without executing them.

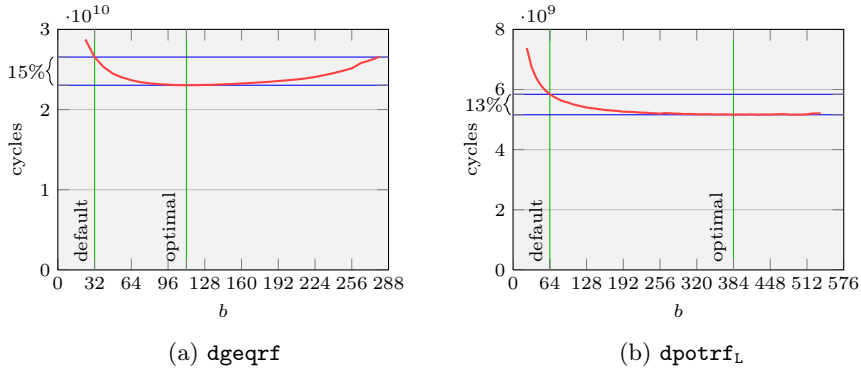
\begin{figure}[t]
    \centering\scriptsize
    \tikzset{external/export=true}

    \begin{subfigure}{.49\textwidth}
        \tikzsetnextfilename{bintro_qr}
        \begin{tikzpicture}
            \begin{axis}[
                twocolplot,
                ymin=0,
                ymax=3e10,
                xmin=0,
                xmax=288,
                xtick={0,32,...,288},
                xlabel={$b$},
                ylabel={cycles}, 
                extra x ticks={32, 112},
                extra x tick labels={,},
                extra x tick style={
                    grid=major,
                    grid style={plot2},
                    tick label style={
                        rotate=90,
                        anchor=west
                    },
                    ticklabel pos=right,
                },
                extra y ticks={26545739031, 23033753712},
                extra y tick labels={,},
                extra y tick style={
                    grid=major,
                    grid style={plot3},
                },
            ]
                \addplot[plot1] file {figures/data/bintro/qr.dat};
                \coordinate (qr def) at (axis cs:-4, 26545739031);
                \coordinate (qr opt) at (axis cs:-4, 23033753712);
                \node[anchor=south west, rotate=90] (x) at (axis cs: 32, 0) {default};
                \node[anchor=south west, rotate=90] (x) at (axis cs: 112, 0) {optimal};
            \end{axis}
            \draw[decorate, decoration=brace] (qr opt) -- (qr def) node[midway, anchor=east] {15\%};
        \end{tikzpicture}
        \caption{\dgeqrf}
        \label{fig:bintro_qr}
    \end{subfigure}
    \begin{subfigure}{.49\textwidth}
        \tikzsetnextfilename{bintro_chol}
        \begin{tikzpicture}
            \begin{axis}[
                twocolplot,
                ymin=0,
                ymax=8e9,
                xmin=0,
                xmax=576,
                xtick={0,64,...,576},
                xlabel={$b$},
                ylabel={cycles}, 
                extra x ticks={64, 384},
                extra x tick labels={,},
                extra x tick style={
                    grid=major,
                    grid style={plot2},
                    tick label style={
                        rotate=90,
                        anchor=west
                    },
                    ticklabel pos=right,
                },
                extra y ticks={5840417745, 5159995515},
                extra y tick labels={,},
                extra y tick style={
                    grid=major,
                    grid style={plot3},
                },
            ]
                \addplot[plot1] file {figures/data/bintro/chol.dat};
                \coordinate (chol def) at (axis cs:-4, 5840417745);
                \coordinate (chol opt) at (axis cs:-4, 5159995515);
                \node[anchor=south west, rotate=90] at (axis cs: 64, 0) {default};
                \node[anchor=south west, rotate=90] at (axis cs: 384, 0) {optimal};
            \end{axis}
            \draw[decorate, decoration=brace] (chol opt) -- (chol def) node[midway, anchor=east] {13\%};
        \end{tikzpicture}
        \caption{\dpotrf}
        \label{fig:bintro_chol}
    \end{subfigure}

    \caption{
        Runtime of \dgeqrf (QR) and \dpotrf (Cholesky) on square matrices of
        size $n = 3{,}800$ with varying block-size $b$.
    }
    
    \label{fig:bintro}
    \tikzset{external/export=false}
\end{figure}

\parsum{optimal block-size}
We motivate here our work by means of a typical scenario.  In order to exploit
BLAS-3 performance, many of LAPACK's routines employ blocked algorithms; these
algorithms work not with scalar and vector matrix portions, but always with
blocks and panels of a predefined {\em block-size} $b$---a crucial optimization
parameter.  In \autoref{fig:bintro}, we illustrate the impact of this block-size
on the performance of LAPACK's QR and Cholesky decompositions (respectively,
\dgeqrf and \dpotrf):  We use one core of an Intel Penryn E5450\footnote{%
    3GHz, 4 cores, 6MB L2 cache per 2 cores, 4 flops/cycle.
} (Harpertown) and link LAPACK to the high-performance {\sc OpenBLAS}
library~\cite{openblas} to decompose of square matrices of size $n = 3{,}800$
with varying block-size $b$.  As the figure shows, optimal performance for
\dgeqrf (QR) is attained with $b = 112$; by contrast, LAPACK's default value is
$b = 32$, for which \dgeqrf is $15\%$ slower.  For \dpotrf (Cholesky), the
message is similar: when using LAPACK's default $b = 64$, the decomposition
takes $13\%$ longer than for the optimal value at $b = 384$.  One of the 
applications of the methodology presented in this paper is to automatically
tuning algorithmic parameters such as these block-sizes without the here
presented empirical experiments.

\parsum{related work}
There exist several works on performance modeling and on the influence of
caching in DLA; we mention here some notable examples.
Cuenca~et~al. developed a system of self-optimizing linear algebra routines
(SOLAR)~\cite{solar}; every routine is associated with performance information,
which is hierarchically propagated to higher level routines in order to tune
them.
Iakymchuk~et~al. model the performance of BLAS analytically based on memory
access patterns~\cite{roman,roman2}; while their models represent the program
execution very accurately, constructing them requires a high level of expertise
of both routines and architecture.
Whaley empirically tunes the block-size for LAPACK routines and emphasizes its
impact on performance~\cite{lapacktuning}.
Lam~et~al. study caching in the context of blocking within DLA
kernels~\cite{blocking1991}.

In contrast, our approach combines highly accurate, measurement based
performance models for compute kernel with an automated analysis of the cache
states present throughout the algorithm execution.  With this combination, we
extend our work on model-based performance predictions~\cite{modeling} from
in-cache problem sizes to the significantly more challenging scenario of
problems not fitting in cache.

\parsum{paper structure}
The rest of this paper is structured as follows:  In \autoref{sec:perfmod}, we
detail our performance modeling technique for DLA kernels.  In
\autoref{sec:cachemod}, we introduce our cache prediction model and its
application in conjunction with performance models to predict DLA algorithms.
In \autoref{sec:results}, we present prediction results and use such predictions
to tune several operations.


    \section{Performance Modeling}
    \label{sec:perfmod}
    In this section we give an overview of our automated performance modeling
framework for dense linear algebra kernels.  Given an architecture and
implementations of the kernel routines (e.g., a BLAS library), this framework,
first introduced in~\cite{modeling}, automatically performs a series of
performance measurements to construct a performance model for each kernel.
Given a set of routine arguments, such a model then yields accurate execution
time estimates.

\subsection{Treatment of Kernel Arguments and Model Structure}
The first step towards constructing performance models of high quality is to
identify the routine features that the models shall capture.  For this purpose,
let us consider the exemplary BLAS kernel \dtrsm---the double precision solve of
a triangular linear system with multiple right hand sides ($B \coloneqq A^{-1}
B$).  This kernel's interface is not only prototypical for other kernels, it
contains all relevant types of arguments:
\begin{center}
    \tt dtrsm(side, uplo, transA, diag, m, n, alpha, A, ldA, B, ldB)
\end{center}

\parsum{argument meaning}
Before studying how these arguments affect the kernel performance, we give a
quick overview of their semantics.
\begin{itemize}
    \item {\tt side}, {\tt uplo}, {\tt transA}, and {\tt diag} are {\em flag
        arguments}.  They identify the precise form of the linear system that is
        solved: The side on which $A$ appears, whether $A$ is lower or upper
        triangular, whether $A$ is transposed, and whether $A$ is
        unit-triangular.
    \item {\tt m} and {\tt n} are {\em size arguments}: $B \in \mathbb
        R^{\mathtt m \times \mathtt n}$ and $A$ correspondingly.
    \item {\tt alpha} is a {\em scalar argument} that
        scales the whole linear system: $B \coloneqq \alpha A^{-1} B$.
    \item {\tt A} and {\tt B} are {\em data arguments} or {\em operands}; they
        point to the first entries of, respectively, $A$ and $B$.
    \item {\tt ldA} and {\tt ldB} are {\em leading dimension arguments}
        corresponding to $A$ and $B$.  While the columns of these matrices are
        stored contiguously (stride 1), the distances of elements of a matrix
        row are given by the leading dimensions (strides {\tt ldA} and {\tt ldB}).
        Distinguishing these row strides from the matrix heights allows to
        operate on sub-matrices, i.e., parts of larger matrices.
\end{itemize}

\parsum{argument effect on performance and their modeling}
We now decide how to represent these arguments in our models by considering
their individual impact on the kernel performance.
\begin{itemize}
    \item {\bf Flag arguments} are limited to few discrete values.  Depending on
        these values, however, the kernel implementation may trigger entirely
        different execution branches with completely independent performance
        characteristics.  To account for these individual characteristics, we
        create a separate performance model for each relevant combination of
        flag arguments (except {\tt diag}).\footnote{%
            in practice, only a limited set of flag combinations is encountered.
        }
    \item {\bf Size arguments} determine the amount of computation and thus
        clearly affect performance.  To account for fine implementation and
        hardware dependent performance characteristics, we used not a single but
        piecewise polynomials to these arguments' impact on execution time.
    \item {\bf Scalar arguments} at first sight do not change the computation
        significantly.  However, some special values---namely $-1$, $0$, and
        $1$---may, trigger separate execution branches in the same way that flag
        arguments do to avoid redundant multiplications.  Hence, we also treat
        these values just like flag arguments, creating a separate model for
        each special value and one for the general case.
    \item {\bf Data arguments}: With few exceptions (such as eigensolvers), the
        executed instructions and thus the performance of dense linear algebra
        kernels do not depend on their operands.  However, performance may
        depend significantly on such arguments' cache locality; kernels will
        execute faster when their operands are in cache.  We reduce the
        generally arbitrarily large set of cache preconditions to two extreme
        cases, where either all operands are {\em in-cache} or they all only
        reside in main memory ({\em out-of-cache}).  Ensuring these cache
        locality conditions while taking measurements, we create two separate
        models for these two cases.
    \item {\bf Leading dimension arguments} change the memory access strides
        when kernels load multiple columns of a matrix simultaneously.  While
        these strides may affect kernel performance slightly for small operands,
        we do not cover their influence in our models and assume large leading
        dimensions, reflecting the common use-case of kernels invocations on
        sub-matrices.
\end{itemize}

\begin{figure}[t]
    \scriptsize
    \centering

    \tikzset{external/export=true}

    \tikzsetnextfilename{modelstructure}
    \newcommand{\sep}{2ex}
    \newcommand{\halfsep}{1ex}
    \newcommand{\doublesep}{4ex}
    \newcommand{\tabheight}{0.6cm}
    \begin{tikzpicture}
        \coordinate (tl) at (0, 0);
        \coordinate (bl) at (0, -6.6);
        \path (tl) ++(\textwidth, 0) coordinate (tr);
        \path (bl) ++(\textwidth, 0) coordinate (br);
        \path (tl) -- (tr) coordinate[midway] (tm);
        \path (bl) -- (br) coordinate[midway] (bm);

        \draw[dotted] (tm) ++(\halfsep, -\tabheight) rectangle (tr) node[midway] {out-of-cache};
        \draw[dotted] (tm) ++(-\halfsep, -\tabheight-\sep) coordinate (corner) -- (corner |- tl) -- (tl) -- (bl) -- (br) -- (tr |- corner) -- cycle;
        \path (corner) ++(0, \sep) -- (tl) node[midway] {in-cache};

        \path (tl) ++(\sep, -\tabheight-\doublesep) coordinate (tl);
        \path (tr) ++(-\sep, -\tabheight-\doublesep) coordinate (tr);
        \path (bl) ++(\sep, \sep) coordinate (bl);
        \path (br) ++(-\sep, \sep) coordinate (br);

        \path (tl) -- (tr) coordinate[pos=.4] (tm);
        \path (tl) -- (tr) coordinate[pos=.8] (tm2);
        
        \filldraw[fill=graybg] (tm) ++(\halfsep, -\tabheight) coordinate(x) (tm2) ++(-\halfsep, 0) rectangle (x) node[midway] {\tt dgemm};
        \filldraw[fill=graybg] (tm2) ++(\halfsep, -\tabheight) rectangle ++(\tabheight, \tabheight);
        \path (tm2) ++(\tabheight+\halfsep, -\tabheight) -- (tr) node[midway] {\ldots};
        \filldraw[fill=graybg] (tm) ++(-\halfsep, -\tabheight-\sep) coordinate (corner) -- (corner |- tl) -- (tl) -- (bl) -- (br) -- (tr |- corner) -- cycle;
        \path (corner) ++(0, \sep) -- (tl) node[midway] {\tt dtrsm};

        \path (tl) ++(\sep, -\tabheight-\sep) coordinate (tl);
        \path (bl) ++(\sep, \sep) coordinate (bl);
        \path (br) ++(-\sep, \sep) coordinate (br);

        \node[anchor=north west] (label) at (tl) {$(\texttt{side}, \texttt{uplo}, \texttt{transA}, \texttt{alpha}) = $};

        \path (label.south west) ++ (0, -\halfsep) coordinate (tl);
        \coordinate (tr) at (tl -| br);
        \path (tl) -- (tr) coordinate[pos=.4] (tm);
        \path (tl) -- (tr) coordinate[pos=.8] (tm2);

        \draw[dashed] (tm) ++(\halfsep, -\tabheight) coordinate(x) (tm2) ++(-\halfsep, 0) rectangle (x) node[midway] {$(\texttt{"L"}, \texttt{"L"}, \texttt{"N"}, 1)$};
        \draw[dashed] (tm2) ++(\halfsep, -\tabheight) rectangle ++(\tabheight, \tabheight);
        \path (tm2) ++(\tabheight+\halfsep, -\tabheight) -- (tr) node[midway] {\ldots};
        \draw[dashed] (tm) ++(-\halfsep, -\tabheight-\sep) coordinate (corner) -- (corner |- tl) -- (tl) -- (bl) -- (br) -- (tr |- corner) -- cycle;
        \path (corner) ++(0, \sep) -- (tl) node[midway] {$(\texttt{"R"}, \texttt{"L"}, \texttt{"N"}, -1)$};

        \path (tl) ++(\sep, -\tabheight-\sep) coordinate (tl);
        \path (bl) ++(\sep, \sep) coordinate (bl);
        \path (br) ++(-\sep, \sep) coordinate (br);

        \node[anchor=north west] (label) at (tl) {$(\texttt{m}, \texttt{n}) \in $};

        \path (label.south west) ++ (0, -\halfsep) coordinate (tl);
        \coordinate (tr) at (tl -| br);
        \path (tl) -- (tr) coordinate[pos=.4] (tm);
        \path (tl) -- (tr) coordinate[pos=.8] (tm2);

        \filldraw[fill=white] (tm) ++(\halfsep, -\tabheight) coordinate(x) (tm2) ++(-\halfsep, 0) rectangle (x) node[midway] {$[24, 4{,}120] \times [24, 536]$};
        \filldraw[fill=white] (tm2) ++(\halfsep, -\tabheight) rectangle ++(\tabheight, \tabheight);
        \path (tm2) ++(\tabheight+\halfsep, -\tabheight) -- (tr) node[midway] {\ldots};
        \filldraw[fill=white] (tm) ++(-\halfsep, -\tabheight-\sep) coordinate (corner) -- (corner |- tl) -- (tl) -- (bl) -- (br) -- (tr |- corner) -- cycle;
        \path (corner) ++(0, \sep) -- (tl) node[midway] {$[24, 536] \times [24, 4{,}120]$};

        \path (tl) ++(\sep, -\tabheight-\doublesep) coordinate (tl);
        \path (tr) ++(-\sep, -\tabheight-\doublesep) coordinate (tr);
        \path (bl) ++(\sep, \sep) coordinate (bl);
        \path (br) ++(-\sep, \sep) coordinate (br);

        \filldraw[fill=plot2, fill opacity=.25] (tl) rectangle (br) coordinate[midway] (mid1);
        \draw (tl) rectangle (bl -| mid1) coordinate[midway] (mid2);
        \draw (tl) rectangle (bl -| mid2) coordinate[midway] (mid3);
        \draw (tl) rectangle (bl -| mid3) coordinate[midway] (mid4);
        \draw (tl |- mid4) rectangle (bl -| mid3) coordinate[midway] (mid5);
        \draw (tl |- mid4) rectangle (bl -| mid5) coordinate[midway] (mid6);

        \path (bm) ++(0, -\sep) node[anchor=north] {
            \fbox{
                \tikz \node[inner sep=\halfsep, outer sep=\halfsep, draw, fill=graybg] (m) {\vphantom{fg}model};
                \hspace{\sep}
                \tikz \node[inner sep=\halfsep, outer sep=\halfsep, draw] (s) {\vphantom{fg}sub-model};
                \hspace{\sep}
                \tikz \node[inner sep=\halfsep, outer sep=\halfsep, draw, fill=plot2!25!white] (s) {polynomial};
            }
        };

    \end{tikzpicture}

    \caption{Structure of the performance models.}
    \label{fig:modelstructure}
    \tikzset{external/export=false}
\end{figure}
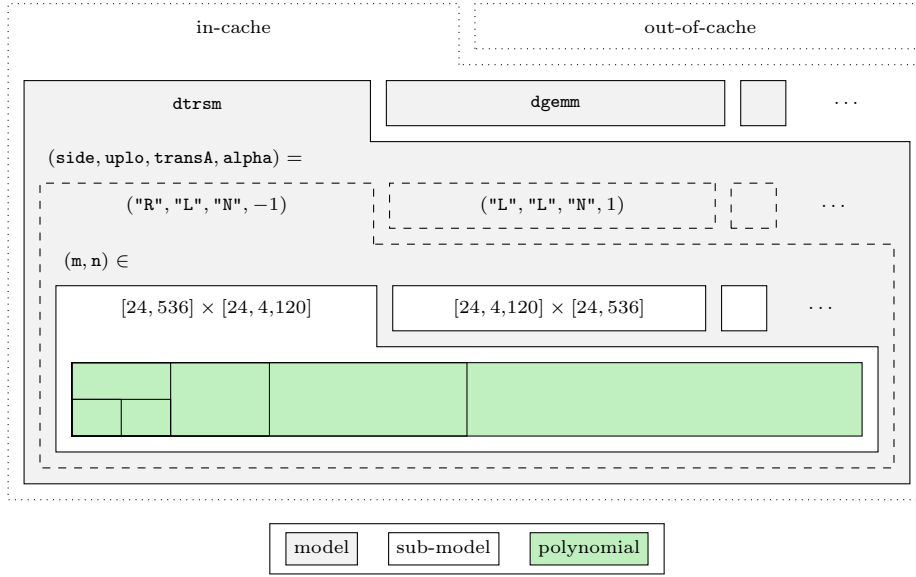

\parsum{summary: model structure}
The structure of our performance models is summarized in
\autoref{fig:modelstructure}:  For each routine we will have two entirely
separate models for in- and out-of-cache situations.  Each of these models is in
turn constructed of a set of {\em sub-models}---one for each combination of flag
and special scalar argument values.  Each sub-model is now only concerned with
modeling execution time as a function of the kernel's size arguments.
Structurally, these models are multivariate piecewise polynomials dividing the
space spanned by the size arguments (which is in practice up to 3-dimensional).

\subsection{Adaptive Refinement}
Let us consider how sub-models are generated for the space spanned by the size
argument(s): starting with a designated rectangular\footnote{%
    In the 2-dimensional case; generally: hyper-cuboidal.
} subspace, (or possibly several distinct sub-spaces), a subdivision is
dynamically obtained through adaptive refinement.  The approach begins by taking
execution time samples on a regular grid spanning the entire considered
subspace; in order to establish a stable basis for the model, this sampling is
repeated multiple times per grid point and the median time is used.  Least
squares fitting is then used to generate a multivariate polynomial which
constitutes a first approximation to the execution time.  In most cases, this
approximation is far too coarse to be acceptable; specifically, if the error
between the least squares fitted polynomial and the samples on the grid is above
a certain threshold, the subspace is subdivided into roughly equally sized
quadrants.  The process of sampling, least squares fitting, and conditionally
subdividing the space is repeated iteratively until a termination criterion
(accuracy or refinement depth) is met.

\begin{figure}[t]
    \centering\scriptsize

    \begin{tikzpicture}[scale=3]
        \begin{scope}[shift={(0, 0)}]
            \filldraw[fill=plot1] (0, 0) rectangle (1, 1);
            \foreach \x in {0.0, 0.1464466094067262, 0.5, 0.8535533905932737, 1.0}
                \foreach \y in {0.0, 0.1464466094067262, 0.5, 0.8535533905932737, 1.0}
                    \shade[ball color=plot3] (\x, \y) circle (.5pt);
            \coordinate (top1) at (.5, 1);
        \end{scope}
        \begin{scope}[shift={(1.25, 0)}]
            \filldraw[fill=plot1!90!plot2] (0,   0) rectangle (.5, .5);
            \filldraw[fill=plot1!60!plot2] (.5,  0) rectangle (1,  .5);
            \filldraw[fill=plot1!20!plot2] (0,  .5) rectangle (.5,  1);
            \filldraw[fill=plot1!10!plot2] (.5, .5) rectangle (1,   1);
            { [scale=.5]
                \foreach \xshift in {0, 1}
                    \foreach \yshift in {0, 1}
                        \foreach \x in {0.0, 0.1464466094067262, 0.5, 0.8535533905932737, 1.0}
                            \foreach \y in {0.0, 0.1464466094067262, 0.5, 0.8535533905932737, 1.0}
                                \shade[shift={(\xshift,\yshift)}, ball color=plot3] (\x, \y) circle (1pt);
            }
            \coordinate (top2) at (.5, 1);
        \end{scope}
        \begin{scope}[shift={(2.5, 0)}]
            \filldraw[fill=plot1!10!plot2] (0,     0) rectangle (.25, .25);
            \filldraw[fill=plot1!40!plot2] (.25,   0) rectangle (.5, .25);
            \filldraw[fill=plot1!30!plot2] (0,   .25) rectangle (.25, .5);
            \filldraw[fill=plot1!00!plot2] (.25, .25) rectangle (.5, .5);

            \filldraw[fill=plot1!10!plot2] (.5,    0) rectangle (.75, .25);
            \filldraw[fill=plot1!20!plot2] (.75,   0) rectangle (1, .25);
            \filldraw[fill=plot1!00!plot2] (.5,  .25) rectangle (.75, .5);
            \filldraw[fill=plot1!00!plot2] (.75, .25) rectangle (1, .5);

            \filldraw[fill=plot1!20!plot2] (0,  .5) rectangle (.5,  1);
            \filldraw[fill=plot1!10!plot2] (.5, .5) rectangle (1,   1);
            { [scale=.25]
                \foreach \xshift in {0, 1, 2, 3}
                    \foreach \yshift in {0, 1}
                        \foreach \x in {0.0, 0.1464466094067262, 0.5, 0.8535533905932737, 1.0}
                            \foreach \y in {0.0, 0.1464466094067262, 0.5, 0.8535533905932737, 1.0}
                                \shade[shift={(\xshift,\yshift)}, ball color=plot3] (\x, \y) circle (2pt);
            }
            \coordinate (top3) at (.5, 1);
        \end{scope}
        \draw[->, shorten >=0.5cm,shorten <=0.5cm] (top1) to[bend left] node[midway, above] {refinement} (top2);
        \draw[->, shorten >=0.5cm,shorten <=0.5cm] (top2) to[bend left] node[midway, above] {refinement} (top3);
    \end{tikzpicture}

    \caption[]{
        Adaptive refinement and sample point selection (\blueball).
    }
    
    \label{fig:adaptiveref}
    \tikzset{external/export=false}
\end{figure}
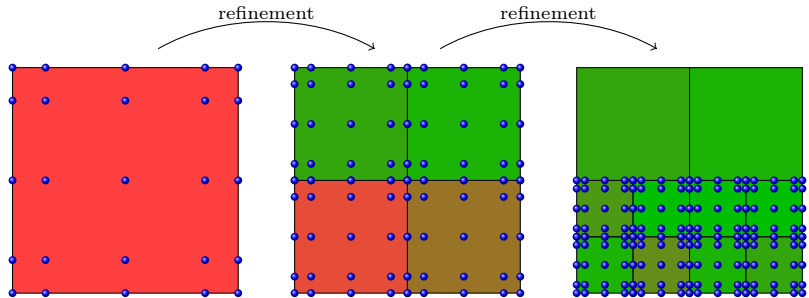

\begin{example}
    An illustration of this adaptive refinement process is given in
    \autoref{fig:adaptiveref} from left to right.  In the beginning, a coarse
    grid of sampling points~(\blueball) yields a first polynomial approximation
    of the routine's performance.  However, since this approximation does not
    satisfy the accuracy requirements (indicated by the color), the space is
    subdivided into equally sized quadrants (\autoref{fig:adaptiveref}, middle).
    In each quadrant a new set of sampling points is chosen and new a new
    polynomial approximation is least squares fitted.  In this example, this
    approximation is already sufficiently accurate for the upper quadrants,
    while the lower quadrants are further subdivided and the refinement is
    recursively repeated.  Subsequent refinement steps could further subdivide
    these resulting quadrants until the refinement's termination criteria are
    reached.
\end{example}

\parsum{modeler configuration options}
Our performance modeling framework has several configuration parameters, which
guide the adaptive refinement based model generation and influence the trade-off
between model accuracy and the number of required samples (and thus the time
spent on generating the models).  In the following we list these parameters
along with the empirical values\footnote{%
    These values assure an excellent model accuracy for a reasonable number of
    samples across various scenarios, such as involving different hardware and
    libraries.
} we use for them in the remainder of this paper:
\begin{itemize}
    \item To avoid small-scale performance fluctuations in the size arguments,
        the coordinates of all sampling points are rounded to multiples of a
        {\em minimum width} of 8.
    \item When sampling a sub-space, we choose the {\em sampling grid} as the
        Cartesian products of Gaussian points along each of its dimensions
        (\blueball in \autoref{fig:adaptiveref}).  Compared to alternatives such
        as a regularly spaced grid or a grid refined on its boundaries, this
        choice has shown to yield the highest accuracy.
    \item The {\em target error bound} for our models (the first refinement
        termination criterion) is chosen at 5\%.  Since this is an upper bound,
        in practice the resulting models will have a significantly higher
        accuracy.
    \item The {\em minimum size} of sub-spaces (the second termination criterion) is
        limited to 32 along each of the space's dimensions.
    \item Our least squares fitter uses polynomials of {\em polynomial degree} 3.
    \item To fit polynomials of degree $n$, one needs at least $n + 1$ points
        along each dimension; however, we {\em oversample} by adding 1 point to
        each dimension for more representative models and more meaningful error
        estimates from these sampling points.
    \item To decide when to refine, the {\em type of error estimate} used to
        evaluate the polynomial fit is the maximum relative error across all
        sampling points.  Alternatives, such as median or average, showed to
        insufficiently capturing highly localized inaccuracies.
\end{itemize}

\subsection{Resulting Sub-space Partitionings}
To illustrate the types of models generated by our approach, we once more
consider \dtrsm; more specifically: sub-models for $\texttt{side} =
\texttt{"L"}$, $\texttt{uplo} = \texttt{"L"}$, $\texttt{transA} = \texttt{"N"}$,
$\texttt{alpha} = 1$, and $\texttt m, \texttt n \in [8, 1024]$.
\autoref{fig:modelplots} shows the partitioning of this 2D domain under
different scenarios.

\input{figures/modelplots}

\parsum{fine and rough}
Figures \ref{fig:modelplots_open_rough} and \ref{fig:modelplots_open_fine},
respectively, show a rough and a fine sub-model for the Intel Penryn running
{\sc OpenBAS}.  For the rough model, we relaxed the adaptive refinement by
lowering the number of grid-points along each axis to 4 (i.e., oversampling
of~0) and chose the average error as the termination criterion.  The fine model
corresponds to the aforementioned parameter choice.  Comparing these two
figures, we see how the refinement further subdivided the regions of
insufficient accuracy, resulting in a better model.

\parsum{different setups}
Next, we apply the fine modeling configuration to different hardware and
software: {\sc ATLAS}~\cite{atlas} is modeled on the Penryn in
\autoref{fig:modelplots_atlas}, while on an Intel Sandy Brdige-EP
E5-2670,\footnote{%
    3.3GHz (Turbo Boost), 8 cores, 20MB shared L3 cache, 8 flops/cycle.
} we used {\sc MKL}~\cite{mkl} in \autoref{fig:modelplots_mkl}.  Although each
model is tiled in a quite different, unique way, all of them cover the
performance characteristics of the corresponding BLAS implementation with
estimation errors mostly well below $2\%$.

\subsection{Trade-off: Accuracy vs. \# Samples}
\begin{figure}[t]
    \centering\scriptsize
    \tikzset{external/export=true}

    \tikzsetnextfilename{modelercfg}
    \begin{tikzpicture}
        \begin{axis}[
            xmin=0,
            xmax=1.3e5,
            ymin=0,
            ymax=1.5,
            xlabel={\#samples},
            ylabel={average error [\%]}, 
            scaled ticks=false
        ]
            \addplot[plot1, only marks] table[x index=1, y expr=\thisrowno{0} * 100] {figures/data/modelercfg/E5450.atlas.dtrsm_.LLN.comp};
            \addlegendentry{ATLAS}
            \addplot[plot2, only marks] table[x index=1, y expr=\thisrowno{0} * 100] {figures/data/modelercfg/E5450.open.dtrsm_.LLN.comp};
            \addlegendentry{OpenBLAS}
            \addplot[plot3, only marks] table[x index=1, y expr=\thisrowno{0} * 100] {figures/data/modelercfg/E5450.mkl.dtrsm_.LLN.comp};
            \addlegendentry{MKL}
            \addplot[black, only marks, mark=o, mark size=4pt] coordinates {
                (30000, 0.495994288918)
                (19425, 0.343973275916) 
                (26850, 0.290401958227) 
            };
            \addlegendentry{selected}
        \end{axis}
    \end{tikzpicture}

    \caption{
        Model accuracy vs. cost (\dtrsm on Penryn).
    }
    
    \label{fig:modelercfg}
    \tikzset{external/export=false}
\end{figure}
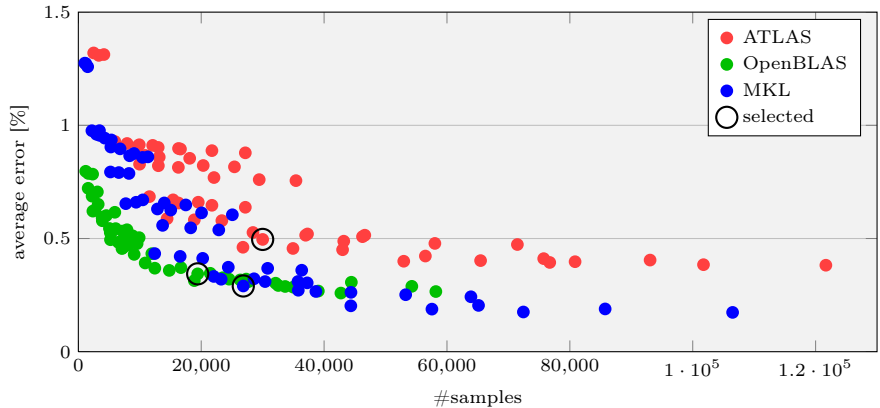

In \autoref{fig:modelercfg}, we consider the three BLAS implementations {\sc
OpenBLAS}, ATLAS, and MKL and alter the adaptive refinement configuration
parameters to highlight the variety of resulting performance models in terms of
their accuracy and cost (i.e., number of samples) to generate them.  The figure
clearly shows the Pareto optimality in the trade-off between accuracy and cost.
For each of these libraries, our framework can generate models of high accuracy
with an average error\footnote{%
    The error was computed by comparing the model estimates with measurements in
    each point of the covered domain that is a multiple of 8 along both {\tt m}
    and {\tt n}.
} below $0.5\%$ and our selected configuration yields a good compromise between
accuracy and model generation costs (\#samples).

All performance measurements suffer from system fluctuations---small variations
in performance resulting from uncontrollable interference from the execution
environment, such as context switches.  On our system, these fluctuations range
from $1.2\%$ for matrices of size $32 \times 32$ to $0.1\%$ for size $1{,}024
\times 1{,}024$---the same order of magnitude as the error in our models, which
can be be as low as $0.2\%$.  While we could strife for even higher accuracies
by statistically accounting for system fluctuations at the cost of considerably
more sample repetitions (cf.~\cite{modeling}), we refrain from doing so in this
paper, since caching proves to have a significantly higher impact on
performance.

\subsection{Cache Unaware Prediction}
\label{sec:unawarepred}
\begin{figure}[t]
    \centering\scriptsize
    \tikzset{external/export=true}

    \tikzsetnextfilename{qricoc}
    \begin{tikzpicture}
        \begin{axis}[
            xmin=0,
            xmax=288,
            ymax=2.2e9,
            xtick={0,32,...,288},
            xlabel={$b$},
            ylabel={cycles}, 
            legend pos=south east
        ]
            \addplot[plot1] file {figures/data/qricoc/oc.dat};
            \addlegendentry{out-of-cache}
            \addplot[plot3] file {figures/data/qricoc/meas.dat};
            \addlegendentry{measured}
            \addplot[plot2] file {figures/data/qricoc/ic.dat};
            \addlegendentry{in-cache}
        \end{axis}
    \end{tikzpicture}

    \caption{
        Prediction of \dgeqrf's runtime from in- and out-of-cache models for
        square matrices of size $n = 1{,}560$ and varying block-size $b$.
    }
    
    \label{fig:qricoc}
    \tikzset{external/export=false}
\end{figure}
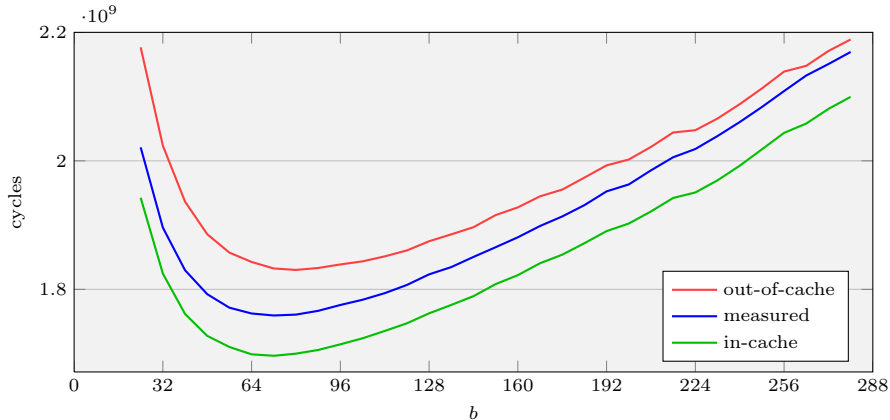

\parsum{QR predictions}
Based on the generated performance models, we can attempt a first prediction of
the performance of LAPACK's QR decomposition \dgeqrf.  For this purpose, we
consider the kernels composing this routine, estimate their runtime using
performance models, and accumulate these runtime estimates into predictions.  In
\autoref{fig:qricoc}, we present these resulting predictions for \dgeqrf on a
square matrix of size $n = 1{,}560$ (1 core of Intel Penryn, {\sc OpenBLAS}) for
both in-cache and out-of-cache
models, alongside measurements of its runtime.  As the figure shows, the
in- and out-of-cache models, respectively, under- and over-estimate the measured
runtime, a phenomenon observed across various scenarios involving different
hardware, BLAS implementations and algorithms.  In the following section, we
introduce a cache tracking methodology that combines these two estimates and
closes this gap.

    \section{Cache Modeling}
    \label{sec:cachemod}
    The framework introduced in the previous section generates highly accurate
performance models for dense linear algebra kernels.  However, we have seen that
this accuracy does not directly translate to accurate performance predictions
for algorithms due to the influence of caching.  In this section, we present a
methodology to obtain high-quality performance predictions; specifically, we use
a cache model that tracks which matrices or sub-matrices are in cache, and,
based on this knowledge, obtain performance predictions through a weighted sum
of the execution time estimates from in-cache and out-of-core models.

\parsum{\dgeqrf}
To illustrate the caching behavior we want to model, \autoref{alg:qr} shows the
exemplary blocked algorithm employed in LAPACK's QR decomposition \dgeqrf.  This
routine decomposes the input matrix $A \in \mathbb R^{m \times n}$ as the
product of an orthogonal matrix $Q \in \mathbb R^{m \times n}$ with an upper
triangular matrix $R \in \mathbb R^{n \times n}$.  Upon termination, $R$ is
stored in $A$'s upper triangular portion, while $A$'s lower portion, together
with an additional output vector $\tau \in \mathbb R^{\min(m, n)}$, represent
$Q$ in the form of elementary reflectors.  In terms of workspace, \dgeqrf
requires an auxiliary buffer $W \in \mathbb R^{m \times b}$, where $b$ is the
algorithmic block-size.  \dgeqrf traverses $A$ in steps of this block-size $b$
diagonally from the top left to the bottom right corner; at each step it
operates on the sub-matrices $A_{11}$, $A_{12}$, $A_{21}$ and $A_{22}$, on
$W_1$, $W_2$, and $\tau$.\footnote{%
    Note that the size of all these operands decreases from one iteration to the
    next.
}

\begin{algorithm}[t]
    \centering\scriptsize
    \newcommand\Aoo{\drawmatrix[height=.33, width=.33, fill=plot1!50]{A_{11}}}
    \newcommand\Aot{\drawmatrix[height=.33, width=.66, fill=plot2!50]{A_{12}}}
    \newcommand\Ato{\drawmatrix[height=.99, width=.33, fill=plot3!50]{A_{21}}}
    \newcommand\Att{\drawmatrix[height=.99, width=.66, fill=plot4!50]{A_{22}}}
    \newcommand\Wo{\drawmatrix[height=.33, width=.33, fill=plot5!50]{W_1}}
    \newcommand\Wt{\drawmatrix[height=.66, width=.33, fill=plot6!50]{W_2}}
    \newcommand\tauo{\drawmatrix[height=.33, width=0, draw=plot7]{\tau_1}}

    \newcommand\AooL{\drawmatrix[height=.33, width=.33, fill=plot1!50, lower]{A_{11}}}
    \newcommand\WoU{\drawmatrix[height=.33, width=.33, fill=plot5!50, upper]{W_1}}
    \newcommand\Aooto{\genfrac{}{}{0pt}{}{\Aoo}{\Ato}{}}

    \begin{minipage}[m]{.49\textwidth}
        \centering
        \begin{tikzpicture}[scale=.024]
            \draw (0, 0) rectangle (127, -159);

            \filldraw[draw=gray, fill=plot1!50] (32, -32) rectangle ++(31, -31) node[black, midway] {$A_{11}$};
            \filldraw[draw=gray, fill=plot2!50] (64, -32) rectangle ++(63, -31) node[black, midway] {$A_{12}$};
            \filldraw[draw=gray, fill=plot3!50] (32, -64) rectangle ++(31, -95) node[black, midway] {$A_{21}$};
            \filldraw[draw=gray, fill=plot4!50] (64, -64) rectangle ++(63, -95) node[black, midway] {$A_{22}$};

            \draw[decorate, decoration=brace] (-4, -159) -- ++(0,   159) node[midway, anchor=east] {$m$};
            \draw[decorate, decoration=brace] (0,     4) -- ++(127,   0) node[midway, anchor=south] {$n$};
            \draw[decorate, decoration=brace] (28,  -63) -- ++(0,    31) node[midway, anchor=east] {$b$};
            \draw[decorate, decoration=brace] (32,  -28) -- ++(31,    0) node[midway, anchor=south] {$b$};

           \draw[very thick, gray, ->, dotted] (32, -16) -- (111, -96);
        \end{tikzpicture}
        \begin{tikzpicture}[scale=.024]
            \draw (0, 0) rectangle (0, -127);

            \filldraw[plot7!50] (0, -32) rectangle ++(0, -31) node[black, midway] {$\tau_1$};
            
            \path (0, 0) -- (0, -159);
        \end{tikzpicture}
        \begin{tikzpicture}[scale=.024]
            \draw[very thick, gray, ->, dotted] (-12, -16) -- (-12, -111);

            \draw (0, 0) rectangle (31, -127);

            \filldraw[draw=gray, fill=plot5!50] (0, 0) rectangle ++(31, -31) node[black, midway] {$W_1$};
            \filldraw[draw=gray, fill=plot6!50] (0, -32) rectangle ++(31, -63) node[black, midway] {$W_2$};

            \draw[decorate, decoration=brace] (35, 0) -- ++(0,  -31) node[midway, anchor=west] {$b$};
            \draw[decorate, decoration=brace] (43, 0) -- ++(0,  -127) node[midway, anchor=west] {$n$};
            \draw[decorate, decoration=brace] (0,  4) -- ++(31,   0) node[midway, anchor=south] {$b$};

            \path (0, 0) -- (0, -159);
        \end{tikzpicture}
    \end{minipage}
    \begin{minipage}[m]{.49\textwidth}
        \centering
        \fcolorbox{black}{graybg}{
            \begin{minipage}{.8\textwidth}
                $\Aooto, \tauo \coloneqq QR\Biggl(\Aooto\Biggr)$\hfill(\dgeqr)\\
                $\WoU          \coloneqq T\Biggl(\Aooto, \tauo\Biggr)$\hfill(\dlarft)\\
                $\Wt           \coloneqq \Aot^T$\hfill($b\times$\dcopy)\\
                $\Wt           \coloneqq \Wt \AooL$\hfill(\dtrmmRLNU)\\
                $\Wt           \coloneqq \Wt + \Att^T \Ato$\hfill(\dgemmTN)\\
                $\Wt           \coloneqq \Wt \WoU$\hfill(\dtrmmRUNN)\\
                $\Att          \coloneqq \Att - \Ato \Wt^T$\hfill(\dgemmNT)\\
                $\Wt           \coloneqq \Wt \AooL^T$\hfill(\dtrmmRLTU)
            \end{minipage}
        }
    \end{minipage}

    \caption{
        QR Decomposition \dgeqrf.  The shapes on the left illustrate \dgeqrf's
        traversal of its data arguments $A$, $\tau$, and $W$.
    }
    \label{alg:qr}
\end{algorithm}
\setcounter{algorithms}\thealgorithm

\parsum{assumptions}
Our goal is to, for each kernel invocation (see \autoref{alg:qr}, right),
determine if the involved sub-matrix operands are available in cache.  For this
purpose, we assume a fully associative Least Recently Used (LRU) cache
replacement policy.\footnote{%
    Due to the regular storage format and memory access strides of dense linear
    algebra operations, this simplifying assumption does not affect the
    reliability of the results.
}  With this policy, determining whether a memory region is in cache or not (or
even in which cache level) boils down to accumulating the size of all memory
regions loaded since and including its last access; if this size, referred to as
the {\em access distance}, is smaller than the cache size, LRU guarantees that
the considered memory regions is in cache.

\parsum{tracking and access distances}
Our first objective is to, within a sequence of compute kernels, identify the
set of memory regions $M$ that were accessed since the last operation involving
a specified kernel operand $\mathcal Op$.  To this end, our approach scans
backwards through the previous kernel invocations, constructing the collection
$M$ of mutually exclusive memory regions from all encountered kernels.  The scan
terminates when either
\begin{itemize}
    \item $\mathcal Op$ is found (all memory regions of the kernel containing
        $\mathcal Op$ are added to $M$),
    \item $M$ is already larger than the cache size, or
    \item the beginning of the kernel sequence is reached (an artificial memory
        region as large as all operands of the entire predicted algorithm is
        added to $M$\footnote{%
            This reflects a repeated execution of the algorithm---the condition
            under which we perform our timings.  Adapting this behavior to other
            scenarios would require knowledge on the surrounding program.
        }).
\end{itemize}
Independently of the termination condition, the resulting collection $M$
contains all memory regions that were accessed since $\mathcal Op$ was last
used.  Computing and summing the sizes of these regions yields the access
distance.

{
\newcommand\Aoo{\drawmatrix[height=.33, width=.33, fill=plot1!50]{A_{11}}}
\newcommand\Aot{\drawmatrix[height=.33, width=.66, fill=plot2!50]{A_{12}}}
\newcommand\Ato{\drawmatrix[height=.99, width=.33, fill=plot3!50]{A_{21}}}
\newcommand\Att{\drawmatrix[height=.99, width=.66, fill=plot4!50]{A_{22}}}
\newcommand\Wo{\drawmatrix[height=.33, width=.33, fill=plot5!50]{W_1}}
\newcommand\Wt{\drawmatrix[height=.66, width=.33, fill=plot6!50]{W_2}}
\newcommand\tauo{\drawmatrix[height=.33, width=0, draw=plot7]{\tau_1}}

\newcommand\AooL{\drawmatrix[height=.33, width=.33, fill=plot1!50, lower]{A_{11}}}
\newcommand\WoU{\drawmatrix[height=.33, width=.33, fill=plot5!50, upper]{W_1}}
\newcommand\Aooto{\genfrac{}{}{0pt}{}{\Aoo}{\Ato}{}}
\begin{example}
    In LAPACK's QR~decomposition (\autoref{alg:qr}), consider the \dtrmmRLNU
    invocation
    $$
        \Wt \coloneqq \Wt \AooL \enspace.
    $$
    For each of the two operands ($W_2$ and $A_{11}$), we now show how the
    access distance is computed by means of a backward scan of the kernels.  
    \begin{itemize}
        \item $\Wt$ was involved in the series of $b$ \dcopy{}s immediately
            preceding the \dtrmmRLNU.  Hence, the scan terminates on condition
            1.~and since the last access to $W_2$ only the operands of these
            \dcopy{}s were loaded into cache:
            $$
                M = \left\{\Wt, \Aot\right\} \enspace.
            $$
        \item $\AooL$ is not involved in the \dcopy{}s, and the involved
            operands are (for reasonable block-sizes) not larger than the cache.
            Therefore the scan goes back to the previous kernel (\dlarft), which
            happens to involve $A_{11}$.  Once again, termination criterion
            1.~is met and the collection $M$ for $A_{11}$ is
            $$
                M = \left\{\Wt, \Aot, \Wo, \Aoo, \Ato, \tauo\right\} \enspace.
            $$
    \end{itemize}
\end{example}

In contrast to this example, our method scans for the last access to a kernel
operand not symbolically as sub-matrices but in the form of memory regions
solely identified by their memory addresses ad sizes.  Hence, scanning works
seamlessly across iterations of blocked algorithms, in each of which the
symbolic sub-matrices refer to different memory regions.

\begin{example}
    The first kernel invocation in each iteration of \dgeqrf is \dgeqr:
    $$
        \Aooto, \tauo \coloneqq QR\Biggl(\Aooto\Biggr) \enspace.
    $$
    \begin{itemize}
        \item From one iteration of the blocked algorithm to the next, the shift
            in the blocking along the diagonal of $A$ (see \autoref{alg:qr},
            left) implies that (unless we are in the first iteration), the
            memory regions now associated with both $\Aoo$ and $\Ato$ were in
            the previous iteration part of $\Att$.  Therefore, an access is
            found in the second last kernel invocation \dgemmNT and (with the
            symbolic shapes referring to the previous iteration) we have
            $$
                M = \left\{\Wt, \AooL, \Att, \Ato\right\} \enspace.
            $$
        \item $\tauo$ on the other hand was never previously accessed, which is
            why the scan terminates on either criterion 2.~(the collection of
            regions exceeds the cache size) or criterion 3.~(the beginning of
            the algorithm is reached).
    \end{itemize}
\end{example}
}

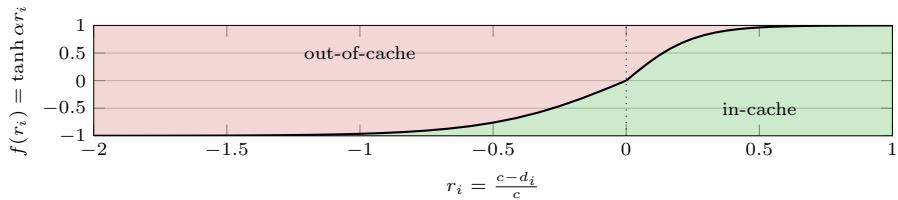
\begin{figure}[t]
    \centering\scriptsize

    \begin{tikzpicture}
        \begin{axis}[
                height=.25\textwidth,
                xlabel={$r_i = \frac{c - d_i}c$},
                ylabel={$f(r_i) = \tanh \alpha r_i$},
                xmin=-2,
                xmax=1,
                ymin=-1,
                ymax=1,
            ]
            \addplot[fill, fill opacity=.15, draw opacity=0, plot2, domain=0:1] {tanh(4 * x)} -- (axis cs:1, -1) -- (axis cs:0, -1) -- cycle;
            \addplot[fill, fill opacity=.15, draw opacity=0, plot1, domain=0:1] {tanh(4 * x)} -- (axis cs:1, 1) -- (axis cs:0, 1) -- cycle;
            \addplot[fill, fill opacity=.15, draw opacity=0, plot2, domain=-2:0] {tanh(2 * x)} -- (axis cs:0, -1) -- (axis cs:-2, -1) -- cycle;
            \addplot[fill, fill opacity=.15, draw opacity=0, plot1, domain=-2:0] {tanh(2 * x)} -- (axis cs:0, 1) -- (axis cs:-2, 1) -- cycle;
            \addplot[domain=0:1] {tanh(4 * x)};
            \addplot[domain=-2:0] {tanh(2 * x)};
            \draw[dotted] (axis cs:0, -1) -- (axis cs:0, 1);
            \node at (axis cs:.5, -.5) {in-cache};
            \node at (axis cs:-1, .5) {out-of-cache};
        \end{axis}
    \end{tikzpicture}

    \caption{
        Smoothing function $f$ for in- and out-of-cache memory regions.
    }
    
    \label{fig:smoothing}
    \tikzset{external/export=false}
\end{figure}

\parsum{weighting an smoothing of ic/ooc}
Once summing up the sizes of the corresponding collections $M$ determines the
access distances for all operands of a kernel, we weight the model estimates for
the in-cache and out-of-cache execution times as follows:  First, we translate
the access distance $d_i$ of each memory region $i$ into a {\em relative access
distance} $r_i$ with respect to the cache size\footnote{%
    We only consider the largest CPU cache, since the data movement between this
    cache and the main memory has the most critical influence on performance.
} $c$:
$$
    r_i = \frac{c - d_i}c \enspace.
$$
In other words, a relative access distance $r_i > 0$ means that region $i$ is
likely still in cache, while $r_i < 0$ means that it is likely no longer in
cache.  To avoid strict classification of in-cache or out-of-cache, we use a
smoothing function $f$ that splits operands with access distances close to the
cache size between these extreme options (see \autoref{fig:smoothing}):
$$
    f(r_i) = \tanh w r_i,
    \text{ where }
    w = \begin{cases}
        4 &\text{ if } r_i \geq 0, \\
        2 &\text{ if } r_i < 0
    \end{cases}
    \enspace.
$$
Here, a value for $f$ of $1$ and $-1$, respectively correspond to entirely
in-cache and out-of-cache,  while intermediate values indicate a distribution
between both scenarios.  Thus, $f(r_i)$ can be understood as a smoothed
association with in- and out-of-cache preconditions for memory region $i$.  From
the $f(r_i)$ of each operand $i$, the cache-state association $\alpha$ for the
entire kernel is obtained by weighting with the corresponding operand sizes
$s_i$:
$$
    \alpha = \frac{\sum_i f(r_i) s_i}{\sum_i s_i} \enspace.
$$
Now, the kernel runtime prediction $p$ is obtained form in- and out-of-cache
model estimates $t_{ic}$ and $t_{oc}$ as
$$
    t = \frac{1 + \alpha}2 t_{ic} + \frac{1 - \alpha}2 t_{oc} \enspace.
$$

\parsum{summary}
As the following section will show, this cache modeling approach yields accurate
performance predictions for entire algorithms from only two cache-aware
performance models for each kernel.


    \section{Results}
    \label{sec:results}
    \subsection{Block-size Optimization}
\parsum{prediction}
In the introduction of this paper, we observed that LAPACK's default block-size
of can be severely suboptimal for algorithms such as the QR decomposition
\dgeqrf and the Cholesky decomposition \dpotrf.  We now employ the methodology
described in Sections \ref{sec:perfmod} and \ref{sec:cachemod} to predict
optimal values for $b$.  Given the size the input matrix, we estimate the
runtime for the considered decomposition algorithm for a range of block-sizes
$b$ and accordingly pick the value that yields the best performance.

\subsubsection{QR: Square Matrices.}
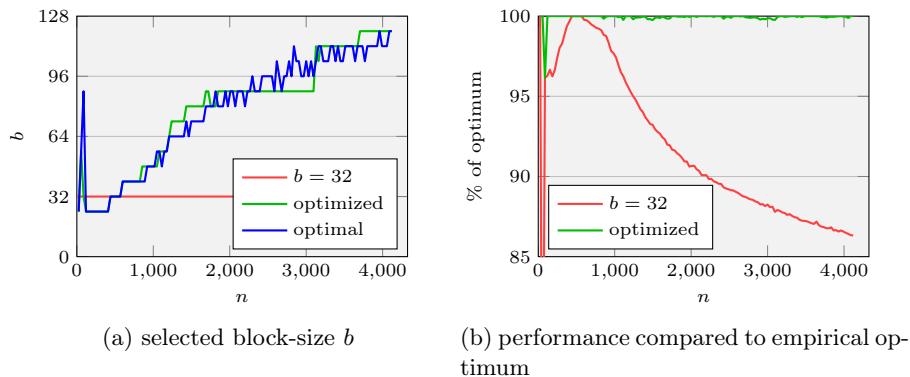
\begin{figure}[t]
    \centering\scriptsize
    \tikzset{external/export=true}

    \begin{subfigure}[t]{.49\textwidth}
        \tikzsetnextfilename{optb_qr_b}
        \begin{tikzpicture}
            \begin{axis}[
                twocolplot,
                ymin=0,
                xmin=0,
                ymax=128,
                ytick={0,32,...,128},
                xlabel={$n$},
                ylabel={$b$}, 
                legend pos=south east,
            ]
                \addplot[plot1, domain=0:4120] {32};
                \addlegendentry{$b = 32$}
                \addplot[plot2] file {figures/data/optb/qr/pred.b};
                \addlegendentry{optimized}
                \addplot[plot3] file {figures/data/optb/qr/opt.b};
                \addlegendentry{optimal}
            \end{axis}
        \end{tikzpicture}
        \caption{selected block-size $b$}
        \label{fig:optb_qr_b}
    \end{subfigure}
    \begin{subfigure}[t]{.49\textwidth}
        \tikzsetnextfilename{optb_qr_eff}
        \begin{tikzpicture}
            \begin{axis}[
                twocolplot,
                ymin=85,
                ymax=100,
                xmin=0,
                xlabel={$n$},
                ylabel={\% of optimum}, 
                legend pos=south west,
            ]
                \addplot[plot1] file {figures/data/optb/qr/32.eff};
                \addlegendentry{$b = 32$}
                \addplot[plot2] file {figures/data/optb/qr/pred.eff};
                \addlegendentry{optimized}
            \end{axis}
        \end{tikzpicture}
        \caption{performance compared to empirical optimum}
        \label{fig:optb_qr_eff}
    \end{subfigure}

    \caption{
        Block-size optimization for \dgeqrf (QR) on square matrices $A \in
        \mathbb R^{n \times n}$; Penryn, {\sc OpenBLAS}, 1 thread.
    }
    
    \label{fig:optb_qr}
    \tikzset{external/export=false}
\end{figure}

To evaluate the quality of our prediction, we consider \dgeqrf on the Intel
Penryn with {\sc OpenBLAS} and square matrices $A \in \mathbb R^{n \times n}$ of
up to size $n = 4{,}120$.  In \autoref{fig:optb_qr}, we compare our
prediction-based estimate for $b$ with both LAPACK's default ($b = 32$) and the
empirically optimal\footnote{%
    The performance measurements for this empirical optimization were very
    time-consuming and took more than 1 day.
} value.
As \autoref{fig:optb_qr_b} shows, the estimates for $b$ closely match the
optimum, even despite the visible fluctuations.
\hyperref[fig:optb_qr_eff]{Figure~\ref*{fig:optb_qr_eff}} presents how the
performance with these optimized $b$ compares to the empirical optimum: With the
exception\footnote{%
    The cause: up to this point, using only the unblocked version (i.e., $b =
    n$) is optimal, while our prediction already suggests the blocked algorithm
    with a small block-size.
} of $n = 88$, our prediction selects the block-sizes so well that we always
attain at least $99.7\%$ of the optimal performance.

\subsubsection{QR: Tall-and-Skinny Matrices.}
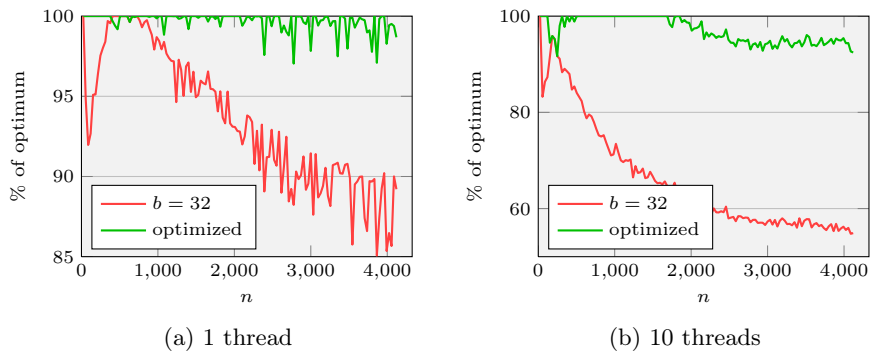
\begin{figure}[t]
    \centering\scriptsize
    \tikzset{external/export=true}

    \begin{subfigure}[t]{.49\textwidth}
        \tikzsetnextfilename{optb_qrts1}
        \begin{tikzpicture}
            \begin{axis}[
                twocolplot,
                ymin=85,
                ymax=100,
                xmin=0,
                xlabel={$n$},
                ylabel={\% of optimum}, 
                legend pos=south west
            ]
                \addplot[plot1] file {figures/data/optb/qrts1/32.eff};
                \addlegendentry{$b = 32$}
                \label{plt:optb_qrts:32}
                \addplot[plot2] file {figures/data/optb/qrts1/pred.eff};
                \addlegendentry{optimized}
                \label{plt:optb_qrts:opt}
            \end{axis}
        \end{tikzpicture}
        \caption{1 thread}
        \label{fig:optb_qrts1}
    \end{subfigure}
    \begin{subfigure}[t]{.49\textwidth}
        \tikzsetnextfilename{optb_qrts10}
        \begin{tikzpicture}
            \begin{axis}[
                twocolplot,
                ymin=50,
                ymax=100,
                xmin=0,
                xlabel={$n$},
                ylabel={\% of optimum}, 
                legend pos=south west
            ]
                \addplot[plot1] file {figures/data/optb/qrts10/32.eff};
                \addlegendentry{$b = 32$}
                \addplot[plot2] file {figures/data/optb/qrts10/pred.eff};
                \addlegendentry{optimized}
            \end{axis}
        \end{tikzpicture}
        \caption{10 threads}
        \label{fig:optb_qrts10}
    \end{subfigure}

    \caption{
        Block-size optimization for \dgeqrf (QR) on tall-and-skinny matrices $A
        \in \mathbb R^{4{,}120 \times n}$; Ivy Bridge-EP, {\sc OpenBLAS}.
    }
    
    \label{fig:optb_qrts}
    \tikzset{external/export=false}
\end{figure}

Next, we consider \dgeqrf for tall-and-skinny matrices $A \in \mathbb R^{4{,}120
\times n}$---a common application of the QR decomposition---on the Intel Ivy
Bridge-EP.  \hyperref[fig:optb_qrts]{Figure~\ref*{fig:optb_qrts}} displays how
\dgeqrf performs with our estimated block-size $b$ compared to the empirically
optimal block-size configuration, using single-threaded and multi-threaded BLAS.
Due to the performance fluctuations from one problem-size to another on this
system (reaching up to $5\%$), which are also clearly visible when using the
default $b = 32$~(\ref*{plt:optb_qrts:32}), our predictions are less accurate
compared to the older Penryn architecture; nevertheless, using 1~thread, our
optimized block-size~(\ref*{plt:optb_qrts:opt}) still reaches around $99\%$ of
the optimal performance.  Using all 10~cores, for large matrices, the efficiency
only decays to $94\%$, and compared to LAPACK's default
block-size~(\ref*{plt:optb_qrts:32}), the algorithm runs $1.6\times$ faster.

\subsubsection{Cholesky.}
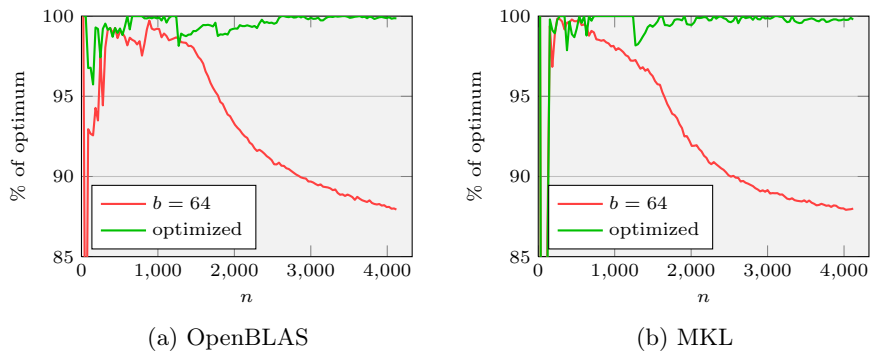
\begin{figure}[t]
    \centering\scriptsize
    \tikzset{external/export=true}

    \begin{subfigure}[t]{.49\textwidth}
        \tikzsetnextfilename{optb_cholopen}
        \begin{tikzpicture}
            \begin{axis}[
                twocolplot,
                ymin=85,
                ymax=100,
                xmin=0,
                xlabel={$n$},
                ylabel={\% of optimum}, 
                legend pos=south west
            ]
                \addplot[plot1] file {figures/data/optb/cholopen/64.eff};
                \addlegendentry{$b = 64$}
                \addplot[plot2] file {figures/data/optb/cholopen/pred.eff};
                \addlegendentry{optimized}
            \end{axis}
        \end{tikzpicture}
        \caption{OpenBLAS}
        \label{fig:optb_cholopen}
    \end{subfigure}
    \begin{subfigure}[t]{.49\textwidth}
        \tikzsetnextfilename{optb_cholmkl}
        \begin{tikzpicture}
            \begin{axis}[
                twocolplot,
                ymin=85,
                ymax=100,
                xmin=0,
                xlabel={$n$},
                ylabel={\% of optimum}, 
                legend pos=south west
            ]
                \addplot[plot1] file {figures/data/optb/cholmkl/64.eff};
                \addlegendentry{$b = 64$}
                \addplot[plot2] file {figures/data/optb/cholmkl/pred.eff};
                \addlegendentry{optimized}
            \end{axis}
        \end{tikzpicture}
        \caption{MKL}
        \label{fig:optb_cholmkl}
    \end{subfigure}

    \caption{
        Block-size optimization for \dpotrf (Cholesky) on $A \in \mathbb R^{n
        \times n}$; Penryn, 1 thread.
    }
    
    \label{fig:optb_chol}
    \tikzset{external/export=false}
\end{figure}

\autoref{fig:optb_chol} contains the results for the block-size optimization for
LAPACK's Cholesky (\dpotrf) on 1~core of the Intel Penryn using {\sc OpenBLAS}
and MKL.  For smaller matrices ($n \leq 3{,}500$), our optimization reaches
$97\%$ of the empirically optimal performance, and for large matrices, we reach
well above $99\%$.  The results are consistent across both BLAS implementations.

\parsum{conclusion}
Together, these experiments provide evidence that our methodology indeed
accomplishes our goal of optimizing the algorithmic block-size across a variety
of different scenarios.

\subsection{Cholesky Algorithm Selection}
For a single mathematical operation, there usually exist several different
algorithms~\cite{Diego,Diego2}.  For such cases, performance predictions can
distinguish the fastest algorithm without executing any of them.  As an example,
we consider the Cholesky decomposition ($L L^T = A$, $A \in \mathbb R^{n \times
n}$ symmetric positive definite).  We compare LAPACK's \dpotrf with the four
algorithms shown in \autoref{algs:chol}: three blocked algorithms and a
recursive one.

\begin{algorithms}[t]
    \centering\scriptsize
    \newcommand\Azz{\drawmatrix[height=.33, width=.33, fill=plot1!50, lower]{A_{00}}}
    \newcommand\Aoz{\drawmatrix[height=.33, width=.33, fill=plot2!50]{A_{10}}}
    \newcommand\Aoo{\drawmatrix[height=.33, width=.33, fill=plot3!50, lower]{A_{11}}}
    \newcommand\Atz{\drawmatrix[height=.66, width=.33, fill=plot4!50]{A_{20}}}
    \newcommand\Ato{\drawmatrix[height=.66, width=.33, fill=plot5!50]{A_{21}}}
    \newcommand\Att{\drawmatrix[height=.66, width=.66, fill=plot6!50, lower]{A_{22}}}

    \newcommand\Atl{\drawmatrix[height=.66, width=.66, fill=plot1!50, lower]{A_{TL}}}
    \newcommand\Abl{\drawmatrix[height=.66, width=.66, fill=plot2!50]{A_{BL}}}
    \newcommand\Abr{\drawmatrix[height=.66, width=.66, fill=plot3!50, lower]{A_{BR}}}

    \begin{minipage}[m]{.39\textwidth}
        \centering
        \begin{tikzpicture}[scale=.03]
            \draw (0, 0) rectangle (127, -127);

            \filldraw[fill=plot1!50, draw=gray] (0,   -0) -- ++(31, -31)  node[black, midway] {$A_{00}$} -| (0, -0);
            \filldraw[fill=plot2!50, draw=gray] (0,  -32) rectangle ++(31, -31) node[black, midway] {$A_{10}$};
            \filldraw[fill=plot3!50, draw=gray] (32, -32) -- ++(31, -31) node[black, midway] {$A_{11}$} -| (32, -32);
            \filldraw[fill=plot4!50, draw=gray] (0,  -64) rectangle ++(31, -63) node[black, midway] {$A_{20}$};
            \filldraw[fill=plot5!50, draw=gray] (32, -64) rectangle ++(31, -63) node[black, midway] {$A_{21}$};
            \filldraw[fill=plot6!50, draw=gray] (64, -64) -- ++(63, -63) node[black, midway] {$A_{22}$} -| (64, -64);

            \draw[decorate, decoration=brace] (-4, -127) -- ++(0,   127) node[midway, anchor=east] {$n$};
            \draw[decorate, decoration=brace] (0,     4) -- ++(127,   0) node[midway, anchor=south] {$n$};
            \draw[decorate, decoration=brace] (68,  -32) -- ++(0,    -31) node[midway, anchor=west] {$b$};
            \draw[decorate, decoration=brace] (32,  -28) -- ++(31,    0) node[midway, anchor=south] {$b$};

            \draw[very thick, gray, ->, dotted] (32, -16) -- (111, -96);
        \end{tikzpicture}
    \end{minipage}
    \begin{minipage}[m]{.59\textwidth}
        \begin{subfigure}[b]{\textwidth}
            \centering
            \fcolorbox{black}{graybg}{
                \begin{minipage}{.8\textwidth}
                    $\Aoz \coloneqq \Aoz \Azz^{-T}$\hfill(\dtrsm)\\
                    $\Aoo \coloneqq \Aoo - \Aoz \Aoz^T$\hfill(\dsyrk)\\
                    $\Aoo \coloneqq \mathrm{chol}(\Aoo)$\hfill(\dpotft)
                \end{minipage}
            }
            \caption{algorithm 1}
        \end{subfigure}

        \vspace{.5\baselineskip}

        \begin{subfigure}{\textwidth}
            \centering
            \fcolorbox{black}{graybg}{
                \begin{minipage}{.8\textwidth}
                    $\Aoo \coloneqq \Aoo - \Aoz \Aoz^T$\hfill(\dsyrk)\\
                    $\Aoo \coloneqq \mathrm{chol}(\Aoo)$\hfill(\dpotft)\\
                    $\Ato \coloneqq \Ato  - \Atz \Aoz^T$\hfill(\dgemm)\\
                    $\Ato \coloneqq \Ato \Aoo^{-T}$\hfill(\dtrsm)
                \end{minipage}
            }
            \caption{algorithm 2 (\dpotrf)}
        \end{subfigure}

        \vspace{.5\baselineskip}

        \begin{subfigure}{\textwidth}
            \centering
            \fcolorbox{black}{graybg}{
                \begin{minipage}{.8\textwidth}
                    $\Aoo \coloneqq \mathrm{chol}(\Aoo)$\hfill(\dpotft)\\
                    $\Ato \coloneqq \Ato \Aoo^{-T}$\hfill(\dtrsm)\\
                    $\Att \coloneqq \Att - \Ato \Ato^T$\hfill(\dsyrk)
                \end{minipage}
            }
            \caption{algorithm 3}
        \end{subfigure}
    \end{minipage}

    \vspace{\baselineskip}

    \begin{minipage}[b]{.39\textwidth}
        \centering
        \begin{tikzpicture}[scale=.03]
            \draw (0, 0) rectangle (127, -127);

            \filldraw[fill=plot1!50, draw=gray] (0,   -0) -- ++(63, -63)  coordinate[midway] (atl) -| (0, -0);
            \filldraw[fill=plot2!50, draw=gray] (0,  -64) rectangle ++(63, -63) node[midway, black] {$A_{BL}$};
            \filldraw[fill=plot3!50, draw=gray] (64, -64) -- ++(63, -63) coordinate[midway] (abr) -| (64, -64);

            \draw[decorate, decoration=brace] (-4, -127) -- ++(0,   127) node[midway, anchor=east] {$n$};
            \draw[decorate, decoration=brace] (0,     4) -- ++(127,   0) node[midway, anchor=south] {$n$};
            \draw[decorate, decoration=brace] (56,  -48) -- ++(0,   -7) node[midway, anchor=west] {$b$};
            \draw[decorate, decoration=brace] (48,  -48) -- ++(7,    0) node[midway, anchor=south] {$b$};

            \node[anchor=north] at (0, -131) {\vphantom{n}};

            \foreach \shift in {0, 64} {
                \begin{scope}[shift={(\shift, -\shift)}, opacity=.7]
                    \draw[gray] (0,   -0) -- ++(31, -31) -| (0, -0);
                    \draw[gray] (0,  -32) rectangle ++(31, -31);
                    \draw[gray] (32, -32) -- ++(31, -31) -| (32, -32);
                \end{scope}
            }

            \foreach \shift in {0, 32, 64, 96} {
                \begin{scope}[shift={(\shift, -\shift)}, opacity=.59]
                    \draw[gray] (0,   -0) -- ++(15, -15) -| (0, -0);
                    \draw[gray] (0,  -16) rectangle ++(15, -15);
                    \draw[gray] (16, -16) -- ++(15, -15) -| (16, -16);
                \end{scope}
            }

            \foreach \shift in {0, 16, 32, 48, 64, 80, 96, 112} {
                \begin{scope}[shift={(\shift, -\shift)}, opacity=.343]
                    \draw[gray] (0, -0) -- ++(7, -7) -| (0, -0);
                    \draw[gray] (0, -8) rectangle ++(7, -7);
                    \draw[gray] (8, -8) -- ++(7, -7) -| (8, -8);
                \end{scope}
            }

            \node at (atl) {$A_{TL}$};
            \node at (abr) {$A_{BR}$};
        \end{tikzpicture}
    \end{minipage}
    \begin{subfigure}[b]{.59\textwidth}
        \centering
        \fcolorbox{black}{graybg}{
            \begin{minipage}{.8\textwidth}
                $\Atl \coloneqq \mathrm{chol}(\Atl)$\hfill(recursion)\\
                $\Abl \coloneqq \Abl\ \Atl^T$\hfill(\dtrsm)\\
                $\Abr \coloneqq \Abr - \Abl\ \Abl^T$\hfill(\dsyrk)\\
                $\Abr \coloneqq \mathrm{chol}(\Abr)$\hfill(recursion)
            \end{minipage}
        }
        \caption{recursive algorithm}
    \end{subfigure}

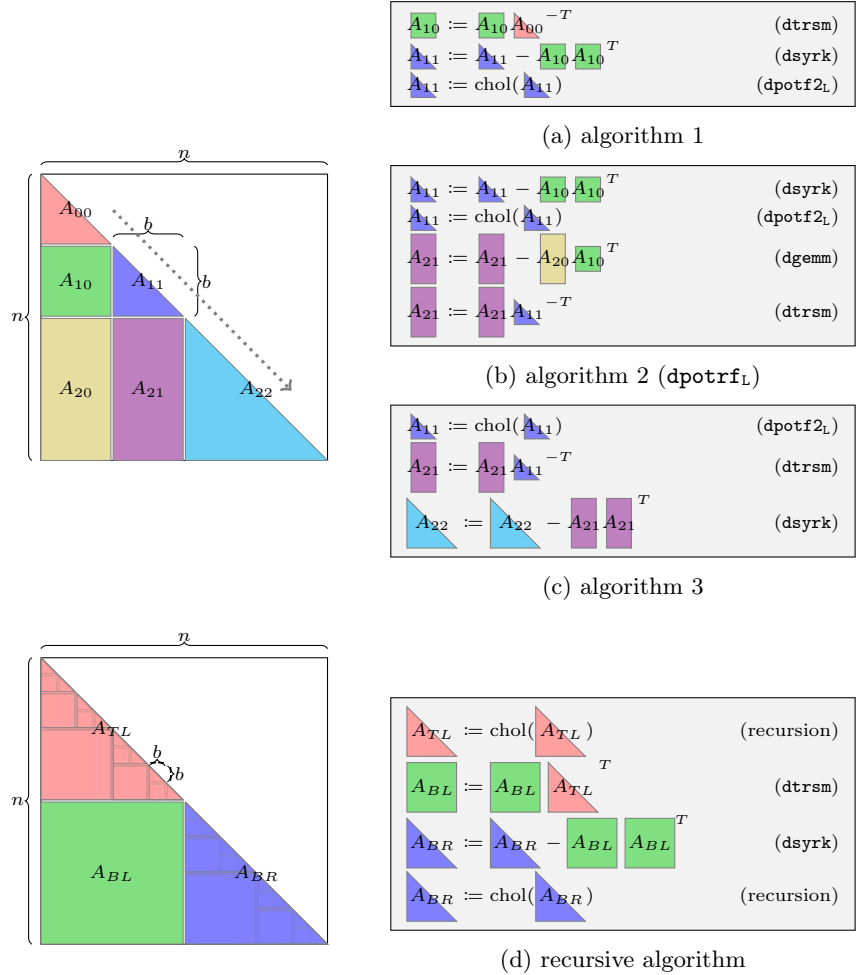
    \caption{Algorithms for Cholesky decomposition.}
    \label{algs:chol}
\end{algorithms}

\parsum{algorithms}
Like \dgeqrf, the blocked algorithms traverse the matrix diagonally from the top
left to the bottom right in steps of the prescribed block-size $b$; for calls to
the unblocked algorithm on $A_{11}$, we use LAPACK's \dpotft.  For these blocked
algorithms, we use a moderate constant block-size of $b = 256$.  The recursive
algorithm cuts the matrix in the middle along both dimensions and recursively
invokes itself for the decompositions of both $A_{TL}$ and $A_{BR}$; only once
the size of these sub-matrices falls below the threshold block-size $b$,
LAPACK's unblocked \dpotft is used.  Preliminary experiments have shown that for
this recursive algorithms, small block-sizes are the best choice irrespective of
the matrix size;\footnote{%
    Small block-sizes do not penalize the performance as for blocked algorithms,
    where they entail a degradation in BLAS-3 performance on small and thin
    operands.
} we therefore use $b = 24$.

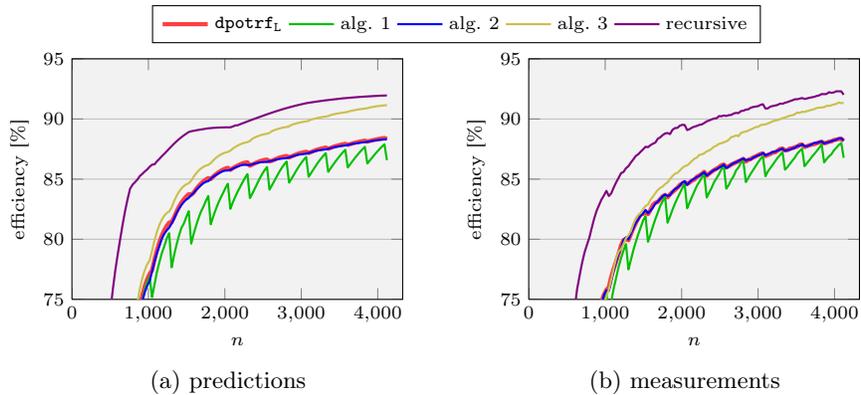
\begin{figure}[t]
    \centering\scriptsize

    \ref*{leg:chol}
    
    \tikzset{external/export=true}

    \begin{subfigure}[t]{.49\textwidth}
        \tikzsetnextfilename{chol_pred}
        \begin{tikzpicture}
            \begin{axis}[
                xmin=0,
                ymin=75,
                ymax=95,
                twocolplot,
                legend to name=leg:chol,
                legend columns=-1,
                xlabel={$n$},
                ylabel={efficiency [\%]},
            ]
                \addplot[plot1, ultra thick] file {figures/data/chol/dpotrf//pred.256eff};
                \addlegendentry{\dpotrf}
                \label{plt:chol:dpotrf}
                \addplot[plot2] file {figures/data/chol/1/pred.256eff};
                \addlegendentry{alg. 1}
                \label{plt:chol:1}
                \addplot[plot3] file {figures/data/chol/2/pred.256eff};
                \addlegendentry{alg. 2}
                \label{plt:chol:2}
                \addplot[plot4] file {figures/data/chol/3/pred.256eff};
                \addlegendentry{alg. 3}
                \label{plt:chol:3}
                \addplot[plot5] file {figures/data/chol/r/pred.24eff};
                \addlegendentry{recursive}
                \label{plt:chol:r}
            \end{axis}
        \end{tikzpicture}
        \caption{predictions}
        \label{fig:chol_pred}
    \end{subfigure}
    \begin{subfigure}[t]{.49\textwidth}
        \tikzsetnextfilename{chol_meas}
        \begin{tikzpicture}
            \begin{axis}[
                xmin=0,
                ymin=75,
                ymax=95,
                twocolplot,
                xlabel={$n$},
                ylabel={efficiency [\%]},
            ]
                \addplot[plot1, ultra thick] file {figures/data/chol/dpotrf//meas.256eff};
                \addplot[plot2] file {figures/data/chol/1/meas.256eff};
                \addplot[plot3] file {figures/data/chol/2/meas.256eff};
                \addplot[plot4] file {figures/data/chol/3/meas.256eff};
                \addplot[plot5] file {figures/data/chol/r/meas.24eff};
            \end{axis}
        \end{tikzpicture}
        \caption{measurements}
        \label{fig:chol_meas}
    \end{subfigure}

    \caption{
        Performance prediction and measurements for four alternative Cholesky
        implementations and \dpotrf.
    }
    
    \label{fig:chol}
    \tikzset{external/export=false}
\end{figure}

\parsum{performance}
Performance predictions and measurements of these Cholesky algorithms using
single-threaded {\sc OpenBLAS} on the Intel Penryn are presented in
\autoref{fig:chol}.  LAPACK's \dpotrf~(\ref*{plt:chol:dpotrf}), which is
identical to algorithm~2~(\ref*{plt:chol:2}), while not the slowest, turns out
to be considerably slower than the fastest blocked
algorithm~3~(\ref*{plt:chol:3}).  Furthermore, the recursive
algorithm~(\ref*{plt:chol:r}) is the fastest by a considerable margin.  Most
importantly, however, our predictions rank the performance of all five
algorithms is correctly.

\subsubsection{Application of the Framework}
Combined into a single framework, our tools and methodology allow the following
workflow:  Given a dense linear algebra routine, such as those in LAPACK, the
framework analyzes the involved compute kernels and, if needed\footnote{%
    Different DLA operations and algorithms often offer a large overlap in terms
    of compute kernels.
} generates or updates the collection of performance models.  With the help of
these models, the routine's tuning parameters, such as the block-size, are then
optimized, reaching within a few percent of the optimal performance.  Moreover,
when facing multiple, possibly automatically generated algorithms for one
operation, our framework can both select the fastest algorithm and optimize it.


    \section{Conclusion}
    In this paper, we presented a highly accurate performance prediction framework
for dense linear algebra algorithms; this framework is based on two components:
performance models for compute kernels and a cache tracking methodology.  In
doing so we extend our previous work~\cite{modeling} on performance model based
predictions to algorithms whose operands exceed the cache size.  

\parsum{performance modeling}
Our modeling framework generates measurement-based performance models for
compute kernels, such as BLAS:  Each routine's argument space is adaptively
subdivided fitting performance samples with polynomial approximations until a
sufficient degree of accuracy is reached.  As a result, we obtain performance
models consisting of piecewise polynomials whose accuracy can match the
magnitude of machine fluctuations.

\parsum{cache modeling}
In order to account for the cache locality of kernel operands, we combine
performance models representing in-cache and out-of-cache situations.  This
combinations is obtained by analyzing the sequence of kernel invocations in the
target algorithm, deducing the cache locality of each of a kernel's operands.

\parsum{results}
This methodology was shown to provide accurate performance predictions for
several dense linear algebra algorithms.  We used these predictions to optimize
the algorithmic block-size for LAPACK algorithms and to select the fastest
member within a family of mathematically equivalent algorithms.


    \section*{Acknowledgments}
    Financial support from the Deutsche Forschungsgemeinschaft (DFG) through
    grant GSC 111 and the Deutsche Telekom Stiftung is gratefully acknowledged.

    \bibliographystyle{splncs}
    \bibliography{references}
\end{document}